\documentstyle[preprint,aps,tighten]{revtex}
\begin{document}
\tighten
\title{Inelastic Electron Lifetime in Disordered Mesoscopic Systems}
\author{Ya.~M.~Blanter$^*$}
\address{Institut f\"ur Theorie der Kondensierten Materie,
Universit\"at Karlsruhe, 76128 Karlsruhe, Germany}
\date{\today}
\maketitle

\begin{abstract}
The inelastic quasiparticle lifetime due to the electron-electron
interaction   
(out-scattering time in the kinetic equation formalism) is calculated
for finite metallic diffusive systems (quantum dots) in the whole
range of parameters.  
Both cases of ``continuous'' (the inelastic level broadening much 
exceeds the mean level spacing) and ``discrete'' spectrum are
analyzed. In particular, crossover between one- and zero-dimensional
regimes is studied in detail. In the case of continuous spectrum the 
out-scattering time is shown to be the same as the inelastic time 
entering expressions for universal conductance fluctuations
and persistent currents. It is also found to be shorter than the 
phase-breaking time in two- and one-dimensional systems, while 
in zero-dimensional systems these two times coincide. In the case of
discrete spectrum for small enough systems a universal behavior of the
scattering time is obtained. For temperatures below the mean level
spacing the out-scattering rate is shown to be vanishingly small.  
\end{abstract}


\bigskip

\section{Introduction}

As is well known, inelastic electron scattering plays an
important role in various phenomena in disordered metallic
systems (see the extensive review by Altshuler and Aronov
\cite{AA}). It is enough to mention that it is responsible for the
weak localization correction to the conductivity. It proved also to be
important in mesoscopic phenomena in these systems. 
In particular, inelastic scattering governs the temperature 
dependence and crossovers between different
dimensionalities for universal conductance fluctuations
(UCF's) \cite{AKm,LSF} and similar problems, such as persistent currents
(see e.g. \cite{SchmidPC}) and correlators of persistent currents 
within the grand canonical ensemble (GCE) (see e.g. \cite{Eckern}). 
Recent estimations \cite{Kamenev} have shown that the conductance of 
mesoscopic systems in the case of a discrete spectrum, at least within the
canonical ensemble (CE), is also governed by the inelastic electron
scattering.  

The electron-electron interaction enters these problems through the 
inelastic scattering time. Extensive investigations carried out about
10 years ago showed that at least two relevant
electron-electron scattering times exist\cite{foot1}: 
(i) the out-scattering time $\tau_{out}$
appearing in the kinetic equation formalism \cite{Schmid,AA-kin}; it
has the meaning of an inverse frequency of inelastic collisions;  
(ii) the phase-breaking time $\tau_{\phi}$ \cite{AAK} (see also 
\cite{Eiler}), responsible, in particular, for the weak localization
correction and for the quasi-particle decay within the Fermi
liquid theory. These two times has been shown to coincide in
three-dimensional systems ($\tau_{\phi} \sim \tau_{out} \propto
T^{-3/2}$). The phase-breaking time is very well studied also in
infinite 2D ($\tau_{\phi} \propto T^{-1}$) and quasi-1D ($\tau_{\phi}
\propto T^{-2/3}$) systems \cite{AAK}. However, the out-scattering rate
in low-dimensional systems $\gamma(\epsilon, T) = \tau_{out}^{-1}$
have caused some controversy. It was studied for zero
temperature by Altshuler and Aronov (see \cite{AA}) who obtained the result  
$\gamma(\epsilon, T = 0) \propto \epsilon^{d/2}$, $d=1$ (Q1D) or $d=2$
(2D). This implies, in particular, that the Fermi liquid theory is
violated in quasi-1D systems 
close enough to the Fermi level. An attempt to include the finite
temperature makes the situation even worse: the out-scattering rate
diverges for low enough energies $\epsilon \ll T$. In two-dimensional
systems attempts to cure this singularity have been made
\cite{AALR,FA,LdS,F,CCKL}, leading to different results.

Below this problem is revisited. The out-scattering rate
$\gamma(\epsilon, T)$ is considered for $T \gg \epsilon$. In
principle, one should 
study the two-parameter problem, but in order not to make the expressions
too cumbersome and not to consider a huge number of parameter ranges,
$\epsilon$ is just put to be equal to zero\cite{foot2}. 
It is shown that the out-scattering time
definitely does not coincide with the phase-breaking one. 
For the 2D case (Sec. \ref{2DD}) just the result by
Refs. \cite{AALR,FA,LdS} $\gamma \propto T\ln T$ is recovered. In the 
quasi-1D system (Sec. \ref{1DD}) new results are obtained: for large
temperatures the 
out-scattering time shows the same temperature dependence  
as the phase-breaking one ($\gamma \propto T^{2/3}$) but is
parametrically shorter than the latter; moreover, for low enough
temperatures (but still in the metallic regime) the out-scattering
rate exceeds the temperature, and exhibits another temperature
dependence $\gamma \propto T^{3/4}$. 

Now we return to the effect of inelastic scattering effect on the 
mesoscopic phenomena in disordered systems, in
particular UCF's. It is rather clear (and it is shown once more below)
that the inelastic scattering time 
entering the UCF problem is essentially the out-scattering time. 
(Note that because of the controversy mentioned above it 
it is believed also to coincide with the phase-breaking time, that is
definitely not the case for 2D and Q1D systems). So one needs to
investigate the out-scattering rate in {\em finite} systems. 
The problem of electron lifetimes in finite systems has not
been addressed so far, except for the only paper by Sivan, Imry and
Aronov \cite{Sivan} (to be discussed below, see Sec. \ref{0DD}).  

The introduction of the finite size of the system leads to the
appearance of two characteristic energies: the Thouless energy $E_c =
D/L^2$ (here $D$ and $L$ are the diffusion coefficient and the largest size
of the system, respectively) and the mean level
spacing $\Delta$. The small parameter of the standard perturbation
theory (in particular, the diagram technique) is $\gamma/\Delta$, and
respectively two cases should be distinguished: ``continuous
spectrum'', $\gamma \gg \Delta$, and ``discrete spectrum'', $\gamma
\ll \Delta$. 

In the three-dimensional systems (e.g. cube of the size $L$) the
situation is rather simple: for $T \gg E_c$ the inelastic level broadening
$\gamma$ much exceeds the mean level spacing $\gamma$ (continuous
spectrum), and the results actually do not depend on $L$. However, for
$T \ll E_c$ one obtains $\gamma \ll \Delta$, and the crossover to the
zero-dimensional situation occurs. So in the case of 3D -- 0D
crossover the distinction between 3D and 0D situations is the
same as one between continuous and discrete spectra (for the
exact formulation of this point see Sec. \ref{0DD}). However, it is not
the case for 2D -- 0D and Q1D -- 0D crossovers. The out-scattering and 
the phase-breaking rates are calculated in Secs. 
\ref{2DD} (2D -- 0D crossover) and \ref{1DD} (Q1D -- 0D crossover)
in the case of the continuous spectrum by means of the diagram
technique. It is shown, in particular, that the spectrum for $T \gg E_c$
is always continuous, and for zero-dimensional situation (the result
depends on $L$) the two inelastic times coincide: $\tau_{out} \sim
\tau_{\phi}$.   

In the case of discrete spectrum $\gamma \ll \Delta$ the perturbation
theory is no longer valid, and the random
matrix theory (RMT) should be involved. Two cases of ``small'' $L \ll
p_Fl^2$ and ``large'' 
$L \gg p_Fl^2$ dots should be distinguished. 
The temperature dependence of the out-scattering
time is obtained $\gamma \sim T^2 \Delta/E_c^2$ for $\Delta \ll T
\ll E_c$ in ``large'' dots, while ``small'' dots exhibit an universal
behavior $\gamma \sim T^2/\epsilon_F$. For the temperatures below the
mean level spacing $T \ll \Delta$ in all cases the scattering rate
$\gamma$ is found to be vanishingly small (Sec. \ref{0DD}). 

Below diffusive metallic systems are considered. ``Diffusive'' implies
the following relation between the characteristic lengths of the
problem: $l \ll L_i$ with $l$ and $L_i$ being the elastic mean free path 
and linear sizes of the system respectively. For definiteness 
a rectangular sample with sizes $L_z \equiv L \ge L_y \ge L_x
\equiv a$ is chosen; $\hbar = 1$. ``Metallic'' means
that (i) the disorder is weak, $p_Fl \gg 1$, $p_F$ being the Fermi
momentum, and (ii) in quasi-1D system the localization length $\xi
\sim l(p_Fa)^2$ is larger than either $L$ or the phase-breaking length
$(D\tau_{\phi})^{1/2}$. The closed systems are of primary interest.
This means, that the coupling to the environment is assumed
to be weak, namely $\tau_{c}^{-1} \ll \Delta$, $\tau_c$ being a
characteristic time of the electron escape through the attached
leads. For open systems the single-particle states are smeared; the
magnitude of this smearing is of order $\gamma_c \sim \min \{E_c,
\tau_c^{-1} \}$, and respectively one is interested in the energy
(temperature) range $\epsilon \gg \gamma_c$. The results obtained
below are valid,
however, for the open systems also provided $\gamma \gg \gamma_c$ (see
the discussion in Sec. \ref{2D}).

\section{Continuous spectrum: general expressions and 2D case} \label{2DD}

\subsection{Generalities}

In the case of continuous spectrum $\gamma \gg \Delta$ the standard 
impurity diagram technique \cite{AGD} may be
involved. We use below the diffusion-cooperon approximation \cite{AA}.
For $T \gg \Delta$ one can not expect any difference between GCE and
CE; as it will be shown below the spectrum is always continuous for $T \gg
E_c \gg \Delta$, so all calculations are performed for the more simple
GCE case.

It is well known that in 3D system the
phase-breaking time 
\begin{equation} \label{tau3D}
\tau_{\phi}^{-1} \sim (T/D)^{3/2}\nu_3^{-1}
\end{equation}
coincides with the out-scattering time and is governed by large
momentum transfer $Dq^2 \sim T$. Here $D = l^2/\tau d$ is the diffusion
coefficient ($d$ is the dimensionality, presently $d=3$) while $\nu_3 =
mp_F/\pi^2$ and $\tau$ are the density of states and the elastic scattering
time respectively. So 3D case is not interesting for us. 
Crossover to other dimensionalities occurs when one of the sizes
is under $(D/T)^{1/2}$. Below low-dimensional systems are studied; the
inequality $T \ll D/a^2$ is assumed to be satisfied.

In the UCF theory and related problems the inelastic scattering time
appears as a result of calculation of the ``UCF diffusion
propagator'' \cite{LSF}. It differs from the ``true'' diffusion 
propagator \cite{AA} by the absence of
electron-electron interaction lines connecting two electron
propagators. A ``true'' density-density correlation function is not
affected by the electron-electron interaction due to the Ward
identity, that is responsible for the cancelation of interaction
lines from the vertex corrections and from the corrections to electron
propagators. However, if some diagrams are omitted, the Ward identity
does not take place any more, and hence the electron-electron
interaction renormalizes the UCF diffusion propagator. In the absence
of magnetic field  the ``UCF cooperon'' is given by exactly the same
expressions as the UCF diffusion propagator.

The diagram equation in the lowest order of the
electron-electron interaction for the diffusion propagator is shown on
Fig.~1 \cite{Schmid,FA}. The shaded rectangle is the full diffusion
propagator $\Gamma(\mbox{\bf q}, \omega)$, the double dashed line is
the bare diffusion propagator 
\begin{equation} \label{gamma_0}
\Gamma^{(0)}(\mbox{\bf q}, \omega) = \frac{1}{\pi\nu_d\tau^2(\vert
\omega \vert + Dq^2)},
\end{equation}
while the crossed rectangle is the block concerned with the
electron-electron interaction $I(\mbox{\bf q}, \omega)$, displayed 
for the UCF case on Fig.~2. The density of states in quasi-$d$
dimensional system $\nu_d$ (to be referred below as $\nu$) is equal to
$\nu_d = \nu_3 a^{3-d}$;
for pure 2D system $\nu_2 = m/\pi$. The Matsubara diagram
technique is used; it is assumed for definiteness that $\omega +
\epsilon > 0$, $\epsilon 
< 0$, and after the analytical continuation $i\omega \to \omega$,
$i\epsilon \to \epsilon$ the propagator corresponds to the averaged
product of retarded and advanced Green's functions 
$\langle G^R(\omega+\epsilon)G^A(\epsilon) \rangle$. The equation can
be readily solved to obtain:
\begin{equation} \label{Gamma}
\Gamma(\mbox{\bf q}, \omega) = \frac{1}{\pi\nu\tau^2 (\vert
\omega \vert + Dq^2 + \gamma)},
\end{equation}
\begin{displaymath}
\gamma = -(\pi\nu\tau^2)^{-1}I(\mbox{\bf q}, \omega)
\end{displaymath}
So the inelastic rate $\gamma$ is given by the interaction block and
is essentially the same as the out-scattering rate in the kinetic equation
approach \cite{Schmid}. 

Now we calculate the interaction block. In the 
arbitrary dimensionality one obtains for the diagrams of Fig. 2a, 2b, 2c
and 2d respectively (note that for Fig.2a and Fig.2b the leading terms
cancel and one has to perform an expansion over the small momenta and
low frequencies):
\begin{displaymath}
I_a = -\pi\nu\tau^2 T \sum_{\Omega > \omega + \epsilon} \sum_{\mbox{\bf
q}_1} \frac{U(\mbox{\bf q}_1,\Omega)}{(\vert \Omega \vert + Dq_1^2)^2}
\left( \omega + \Omega + D(q^2 + q_1^2) \right)
\end{displaymath}
\begin{equation} \label{arbdim}
I_b =  -\pi\nu\tau^2 T \sum_{\Omega < \epsilon} \sum_{\mbox{\bf
q}_1} \frac{U(\mbox{\bf q}_1,\Omega)}{(\vert \Omega \vert + Dq_1^2)^2}
\left( \omega - \Omega + D(q^2 + q_1^2) \right)   
\end{equation}
\begin{displaymath}
I_c =  \pi\nu\tau^2 T \sum_{\Omega < \epsilon + \omega} \sum_{\mbox{\bf
q}_1} \frac{U(\mbox{\bf q}_1,\Omega)}{\omega - \Omega + D(\mbox{\bf q}
- \mbox{\bf q}_1)^2}
\end{displaymath}  
\begin{displaymath}
I_d =  \pi\nu\tau^2 T \sum_{\Omega > \epsilon} \sum_{\mbox{\bf
q}_1} \frac{U(\mbox{\bf q}_1,\Omega)}{\omega + \Omega + D(\mbox{\bf q}
+ \mbox{\bf q}_1)^2}
\end{displaymath} 
Here $U(\mbox{\bf q},\Omega)$ is a screened Coulomb interaction. 

\subsection{2D case: Infinite system}

As the Coulomb interaction has different forms in different
dimensionalities \cite{AA}, the  
analytical continuation should be performed separately for $d=2$ 
and $d=1$. We start from the 2D case with
\begin{equation} \label{U,d=2}
U(\mbox{\bf q},\Omega) = \frac{2\pi e^2}{q} \frac{\vert \Omega \vert +
Dq^2}{\vert \Omega \vert + D \kappa q}
\end{equation}
Here $\kappa = 4\pi e^2 \nu_3a$ for quasi-2D case and $\kappa =
2\pi e^2 \nu_2$ for 2D case. 

After the analytical continuation $i\omega \to \omega$, $i\epsilon \to
\epsilon$ one obtains for $\omega = 0$, $\mbox{\bf q} = 0$:
\begin{equation} \label{nesamosogl}
\gamma = \frac{e^2}{\pi} \int_0^{\infty} D^2 \kappa q^3
dq \int dz \left( \coth \frac{z}{2T}  - \tanh \frac{z-\epsilon}{2T}
\right) \frac{z}{(z^2 + D^2 q^4)(z^2 + D^2 \kappa^2 q^2)}
\end{equation}
It is important that the terms with the hyperbolic cotangent come from
the quantities $I_c$ and $I_d$. 

It is easy to see that for non-zero temperature the integral diverges
at $z=0$. This singularity can be cured by the substitution of
self-consistent expression for the diffusion propagator (\ref{Gamma})
to the formulae (\ref{arbdim}) instead of $\Gamma^{(0)}$. The diffusion
propagators entering the diagrams of Fig.~2a and 2b are ``true'' and
are not renormalized by the electron-electron interaction, while those
of diagrams Fig.~2c and 2d are renormalized. For $T \gg \epsilon$ the main
contribution to the integral comes from the region $\vert z \vert <
2T$, and one obtains the self-consistent equation for
$\gamma$ (that is supposed to be momentum-independent for a while):  
\begin{equation} \label{tau2d-int}
\gamma = \frac{4e^2T}{\pi} \int_{0}^{\infty} D\kappa q_1
\left( D q_1^2 + \gamma \right) dq_1
\int_{0}^{2T} \frac{dz}{(z^2+D^2 \kappa^2 q_1^2) [z^2 + (Dq_1^2 +
\gamma)^{2}]} 
\end{equation}
A straightforward calculation leads to an equation
\begin{equation} \label{tau2d-res}
\gamma = \frac{e^2T}{D \kappa} \ln (2 D \kappa^2 T
\gamma^{-2}) = \frac{e^2T}{D \kappa} \ln \left( 2 D \kappa^2 \frac{(D
\kappa)^2}{e^4 T} \right)
\end{equation}
where the last equality is the leading term of the solution.
Substituting the explicit expressions for $\kappa$, one finally obtains:
\begin{eqnarray} \label{2D}
\gamma = \left\{ \begin{array}{lr}
\left (T/2\epsilon_F\tau \right) \ln \left( 8 D \kappa^2 (\epsilon_F
\tau)^2/T \right) & \mbox{2D} \\ \left(3\pi T/4p_F^2 a l \right) \ln
\left( D \kappa^2 (4 p_F^2 a l/3 \pi)^2 T^{-1} \right) &
\mbox{Quasi-2D} \end{array} 
\right.
\end{eqnarray}
Note that this rate is higher (by a logarithmic factor) than the 2D
phase-breaking rate \cite{AAK}
\begin{eqnarray} \label{2dfi}
\tau_{\phi}^{-1} \sim (T/D \nu_2) \left\lbrace \begin{array}{lr}
\ln (p_F l) & \mbox{2D} \\
(p_Fa)^{-1} \ln (p_F^2 l a) & \mbox{Quasi-2D} 
\end{array} \right.
\end{eqnarray}
The result (\ref{2D}) is valid under the condition $T \gg \gamma$
only. In principle, this condition (unlike the condition of the
Fermi-liquid theory validity $T\tau_{\phi} \gg 1$) is violated in the
exponentially small range of low temperatures. We consider this
problem as purely academic, however, and do not write down
expressions for the corresponding range.

The expression (\ref{2D}) was derived first in
Ref. \cite{AALR} as an imaginary part of the impurity-averaged 
self-energy. Later this
result was confirmed in Refs. \cite{FA,LdS} (Fukuyama and Abrahams \cite{FA}
have used the technique identical to that of the present
paper). However it was believed to be the expression for the
phase-breaking time in a 2D system. The origin of this confusion lies
in the statement\cite{FA} that the diagrams with the interaction 
between different electron lines (i.e.those not contributing to
$\gamma$, but improtant for $\tau_{\phi}$, in-scattering terms in the
kinetic equation approach\cite{Schmid}) are small. However, the
original calculation of the phase-breaking time\cite{AAK} allows one
to separate these two contributions (with and without interaction
lines between different Green's functions) explicitly, and the result
is that these contributions are of the same order. So the difference
between $\gamma$ and $\tau_{\phi}^{-1}$ should be looked for in the
diagrams omitted in Ref. \cite{FA}. Hence the attempt\cite{F} 
to cure this discrepancy by the introduction of a small-momenta cutoff
does not seem well justified. Indeed, the momenta range proposed to be
cut off is exactly the region of relevant momenta in the integral, and
the introduction of finite momenta does not change the result. So we
argue that the result by Fukuyama and Abrahams, (\ref{2D}), is true;
however, it describes not the  
phase-breaking time, but the out-scattering time (or the UCF inelastic
time). We have shown that the inelastic time entering
the UCF problem is essentially the out-scattering time and differs
from the phase-breaking one. It will be seen in Sec. \ref{1DD} that
the difference is even more pronounced for the quasi-one-dimensional
case.  

\subsection{2D case: Finite system}

Now we turn to the description of 2D -- 0D crossover. 
In the derivation of Eq. (\ref{2D}) it was supposed implicitly that
the system is infinite. Let us now consider 2D finite system. Under
the conditions of diffusive regime $L \gg l$, and in the case $T \gg
E_c$ the equation (\ref{tau2d-int}) is still valid, however the
integral should be understood as a sum over momenta with the boundary
conditions taken into account. So $\mbox{\bf q} =
(\pi/L)(n_x,n_y)$ with integers $n_x$ and $n_y$ allowed or not
to be equal to zero, subject to the boundary conditions. It is
important that in any case the $\mbox{\bf q}=0$ mode does not contribute
to the integral: for open systems (attached to the
metallic junctions, i.e. in the UCF problem) this mode is forbidden by
the boundary conditions, while for closed ones (e.g. in the problem
of persistent currents) the conditions of the charge neutrality
imply $U(0,\omega) = 0$. So the difference between these two 
cases is not quite important, and the boundary conditions will be
taken into account just by the cutting off the integral for small
momenta $q < \pi/L$. 

The analysis of the equation (\ref{tau2d-int}) then shows 
that for $\gamma \gg (E_cD\kappa^2)^{1/2}$ this cutoff does not play
any role and the result 
(\ref{2D}) is still valid. Note that this region corresponds to the 2D
case from the diffusion point of view. 
For the opposite case, that is realized under the
conditions
\begin{eqnarray*}
T/\epsilon_F \ll \left\{ \begin{array}{lr} (l/L)\epsilon_F\tau, &
\mbox{2D} \\ 
(l/L) (p_F^2al), & \mbox{Quasi-2D} \end{array} \right.      
\end{eqnarray*}
one obtains 
\begin{equation} \label{2D-0D} 
\gamma = \frac{e^2T}{\kappa D} \ln \frac{2T}{\pi^2 E_c},
\end{equation}
It is seen that this result also differs from Eq. (\ref{tau2d-res}) 
by a logarithmic factor.

\section{Quasi-1D case and Q1D--0D crossover} \label{1DD}

Let us now turn to the quasi-1D case. The screened Coulomb interaction is 
\begin{equation} \label{U,d=1}
U(\mbox{\bf q},\Omega) = e^2 \ln \left(\frac{1}{q^2a^2} \right)  
\frac{\vert \Omega \vert + Dq^2}{\vert \Omega \vert + D q^2 + e^2 \nu_1
Dq^2 \ln (q^2a^2)^{-1}}
\end{equation} 
It is convenient to denote the denominator of this expression as
$\vert \Omega \vert + Dq^2f(q)$ with $f(q) = 1 + C \ln (q^2a^2)^{-1}$,
$C = e^2\nu_1$. If, as usual, $e^2 \sim v$, the constant $C$ occurs to
be of order $C \sim (p_Fa)^2 \gg 1$. 

Prior to the calculation of $\gamma$ for the quasi-1D system it is
necessary to analyze the expression for the phase-breaking time
\cite{AAK} that can be conveniently rewritten as
\begin{equation} \label{taufi-1D}  
\tau_{\phi}^{-1} \sim T^{2/3}\tau^{-1/3} (p_Fa)^{-4/3}
\end{equation}
This result was obtained under some assumptions to be analyzed here. 
First, the Fermi liquid theory is valid only in the case $T\tau_{\phi} \gg
1$. Second, the system is one-dimensional, i.e. $E_c \sim D/L^2 \ll
\tau_{\phi}^{-1} \ll T \ll D/a^2$. Finally, it is
supposed to be metallic, i.e the correlation length $\xi = l (p_Fa)^2
\gg \min \{ L, (D\tau_{\phi})^{1/2} \}$. 

The last condition is violated in the localized regime (the shaded range
on Fig.~3): 
$$T\tau \ll (p_Fa)^{-4}, \ \ l/L \ll (p_Fa)^{-2},$$
and the first three are consequently summarized as (Fig.~3, ranges I,
II, III, and IV):
\begin{equation} \label{param}
(l/L)^3(p_Fa)^2 \ll T\tau \ll (l/a)^2 \ll 1
\end{equation}

The question is of much interest what happens if the condition
$T\tau \gg (l/L)^3(p_Fa)^2$ (for $l/L \gg (p_Fa)^{-2}$) is
violated. The expression for the 
phase-breaking time in an infinite system (\ref{taufi-1D}) is no more
valid there. However, it is possible to derive the expressions for this
regime. This is done in the Appendix A. The result is
\begin{equation} \label{dephas}
\frac{1}{\tau_{\phi}} \sim \frac{T}{(p_Fa)^2} \frac{L}{l}, 
\end{equation}
\begin{equation} \label{param1}
(l/L)^2 \ll T\tau \ll (l/L)^3(p_Fa)^2
\end{equation}
This expression is valid until $E_c \sim T$ (range V), as is seen from
Eq. (\ref{param1}). As will be shown, in the whole range
(\ref{param1}) the inequality $\gamma \gg \Delta$ is
satisfied, $\Delta = (\nu_3V)^{-1}$ being the mean level spacing. So
the spectrum is continuous and this range is subject to the 
perturbative analysis. Thus, one has a ``true'' quasi-1D region
(\ref{param}), transition Q1D -- 0D region (\ref{param1}) and ``true''
0D region $T \ll 
E_c$ where the spectrum is discrete. In this Section both 
Q1D and transitional ranges are considered. It will be shown that a
new splitting of the range in respect to the out-scattering time appears.

Now it is possible to start the calculation of the out-scattering
time. Consideration of the diagrams Fig.~2 leads to the divergent
expression, and this divergence can be cured exactly in the
same way as in 2D case. Extracting the main term for $T \gg \epsilon$,
one obtains:
\begin{equation} \label{eq1D}
\gamma = \frac{4Te^2C}{\pi^2} \int_{\pi/L}^{\infty} dq Dq^2
(Dq^2 + \gamma) \ln^2 (q^2a^2)^{-1} \int_0^{2T} dz \frac{1}{[z^2 +
D^2q^4f^2(q)][z^2 + (Dq^2 + \gamma)^2]}
\end{equation}
Here a low-momentum cutoff is introduced in the momentum integral (see the
discussion for 2D case). Equation (\ref{eq1D}) contains all
information about the scattering rate and should be solved in
different limiting cases (Fig.~3).

\medskip

\noindent {\bf A} (Range I). Let us assume $D(\pi/L)^2f(\pi/L) \ll
\gamma \ll T$. In this case the cutoff is not important, and the  
essential momenta in the $q$-integration are $Dq^2f(q) < \gamma$. The 
equation simplifies to a form:
\begin{displaymath}
\gamma = 4Te^2/CDq_0\pi, \ \ \ \ \ Dq_0^2f(q_0) = \gamma
\end{displaymath}
(the inequality $C \gg 1$ has been used). This yields the result:
\begin{equation} \label{1D}
\gamma = \left( \frac{4Te^2}{\pi} \right)^{2/3} \left( CD
\right)^{-1/3}  \ln^{1/3} \left[ \frac{(CD)^{4/3}}{a^2} \left(
\frac{\pi}{4Te^2} \right)^{2/3} \right]
\end{equation}
This cumbersome expression can be simplified if one assumes again $e^2 \sim
v$. Then, as the cube root of the logarithm is always a quantity of
order unity, one obtains:
\begin{equation} \label{1D-n}
\gamma \sim T^{2/3}\tau^{-1/3}(p_Fa)^{-2/3}
\end{equation}
It is seen that the initial assumptions $D(\pi/L)^2f(\pi/L) \ll
\gamma$ and $\gamma \ll T$ are satisfied only in the temperature range
$$\max \{ (p_Fa)^{-2},(l/L)^3(p_Fa)^4 \} \ll T\tau \ll (l/a)^{2}$$
(Range I). In particular, this result is valid for the infinite
quasi-1D system for temperatures above $\tau^{-1}(p_Fa)^{-2}$. 
It should be emphasized also that in this case $\tau_{\phi} \gamma =
(p_Fl)^{2/3}  
\gg 1$, and consequently $\tau_{out} = \gamma^{-1} \ll \tau_{\phi}$. 
So the out-scattering time in this range of
parameters shows the same temperature dependence as the phase-breaking
time, but is much shorter than the latter. Nevertheless, the
transition to the 0D behavior in the problem of out-scattering times
occurs before the corresponding transition in the dephasing problem.

\medskip

\noindent {\bf B} (Ranges II, III). If one assumes  $D(\pi/L)^2f(\pi/L)
\ll \gamma$, 
$\gamma \gg T$, the straightforward calculation leads to the equation
as follows:
\begin{equation} \label{wz1}
\gamma = \frac{4Te^2}{\pi^2} \int_{\pi/L}^{\infty} dq \arctan
\frac{2T}{Dq^2 f(q)} \frac{1}{Dq^2 + \gamma}
\end{equation}
A new relevant length scale appears, $\tilde q$, defined by $D\tilde q
f(\tilde q) = 2T$. Then two limiting cases should be distinguished: 

\medskip

\noindent 1. (Range II). $\tilde q \gg L^{-1}$. The cut-off is again
unimportant, and one obtains:
\begin{equation} \label{wz2}
\gamma = \frac{2eT^{3/4}}{\pi^{1/2}(CD)^{1/4}} \ln^{-1/4} \left(
\frac{CD}{2Ta^2} \right) \sim T^{3/4} \tau^{-1/4} (p_Fa)^{-1/2}
\end{equation}

The second identity is the estimation, based upon assumption $e^2 \sim
v$. All assumptions made for the derivation of Eq. (\ref{wz2}) occur
to be consistent in the range II: 
$$\max \{ (p_Fa)^{-4}, (l/L)^2(p_Fa)^2 \} \ll T\tau \ll (p_Fa)^{-2}$$

This result does not contain the length of the sample again and
consequently describes the infinite quasi-1D systems for the
temperatures just above the metal-insulator transition: $(p_Fa)^{-4}
\ll T\tau \ll (p_Fa)^{-2}$. We have obtained a new temperature
dependence, $\gamma \propto T^{3/4}$, and the out-scattering time is
now not only parametrically shorter than the phase-breaking one, but
even $\gamma \gg T$. As $\gamma$ has a meaning of a frequency of
electron-electron collisions, the violation of the condition $\gamma
\ll T$ has nothing to do with the violation of Fermi liquid theory
(the latter is subject to another condition, $T\tau_{\phi} \gg 1$).   

\medskip

\noindent 2. (Range III). $\tilde q \ll L^{-1}$. Now the scattering rate
introduced to the rhs of Eq. (\ref{wz1}) for self-consistency, is
inessential, and one obtains
\begin{equation} \label{wz3}
\gamma = \frac{4T}{\pi^{3/2}} \left( \frac{e^2 L}{CD} \right)^{1/2}
\ln^{-1/2} \frac{L}{\pi a} \sim T (l/L)^{1/2} (p_Fa)^{-1}
\end{equation}
This result is valid under conditions (range III):
$$l/L \ll (p_Fa)^{-2}, \ \ \ (p_Fa)^{-4} \ll T\tau \ll
(l/L)^2(p_Fa)^{-2}$$
Now the result contains the length of the sample, but still one has
$\gamma \gg T$ and $\gamma \tau_{\phi} \gg 1$.

\medskip

\noindent {\bf C} (Ranges IV + V). 
In the case $D(\pi/L)^2f(\pi/L) \gg \gamma$ the important region
in the integral over $z$ is $z < Dq^2 + \gamma$, and again
straightforwardly all dependence on the inelastic time cancels out in the
rhs. The result is 
\begin{equation} \label{1D-0D}
\gamma = \frac{2T}{(p_Fa)^2} \frac{L}{l} 
\end{equation}
$$(l/L)^2 \ll T\tau \ll (l/L)^3(p_Fa)^4$$

To summarize, we have described the crossover between
quasi-one-dimensional and zero-dimensional behavior for the case of
continuous spectrum. We have discovered 5 different parameter ranges
(Fig.~3), that can be divided into three groups:
\begin{description}

\item[1.] I and II. Here both the dephasing and the out-scattering are purely
quasi-one-dimensional, the results do not contain $L$. In the whole
range the relation $\tau_{out} \ll \tau_{\phi}$ holds.

\item[2.] III and IV. The transitional region: the dephasing is still  
quasi-one-dimensional, while the out-scattering time depends on $L$
and is of zero-dimensional nature. Still one has $\tau_{out} \ll
\tau_{\phi}$.  

\item[3.] The range V is truly zero dimensional, the expressions for
the phase-breaking rate (\ref{dephas}) and for the out-scattering rate
(\ref{1D-0D}) coincide. However the spectrum still is continuous,
i.e. $\tau_{\phi}^{-1} \sim \tau_{out}^{-1} \gg \Delta$. 

\end{description}

The spectrum becomes discrete for the temperatures below the Thouless
energy: $T \ll E_c$, as it takes place in 3D case.

\section{Discrete spectrum} \label{0DD}

The diagram technique used in Secs. \ref{2DD}, \ref{1DD} is valid only
in the case of continuous
spectrum $\gamma \gg \Delta$. As was shown above, this condition in
all cases is equivalent to another one $T \gg E_c$. So this
perturbative approach is not
suitable for the case of low temperatures (or, alternatively, small
dots) $T \ll E_c$. Fortunately, this is exactly the range where 
the random matrix theory (RMT)
\cite{matr} works quite well for the description of small
disordered systems \cite{GE}. Note that usually
RMT in the range of its applicability is
equivalent to the zero-dimensional non-linear sigma model
\cite{Efetov}. However, in the problem under consideration one should use
the four-point correlation function which up to now was not derived by
means of the supersymmetry method. So RMT (in spirit of the paper by 
Gor'kov and Eliashberg \cite{GE}) seems presently to be the only method
for investigation of the electron-electron interaction in the
non-perturbative regime. To avoid misunderstanding, we stress that the
electron-electron interaction is taken into account {\em
perturbatively}; however, the parameter $\Delta/\gamma$ is no more
small and should be treated {\em non-perturbatively}.

In order not to overburden our expressions, we consider in this
Section the cubic sample with size $L$. The system is assumed to be
charge neutral, and the results rely heavily on this fact. In
principle, the methods used
allow one to consider the samples of arbitrary geometry, and all
expressions below containing the parameters $\Delta$ and $E_c$ instead
of $L$ are valid for the general case (for the
metallic regime). Also, we do not distinguish two types of the
scattering rates from each other. To understand the results better, we
extend the ranges of parameter to the clean case also. The exact
result for the scattering rate in bulk 3D case reads as \cite{Schmid}
\begin{equation} \label{3Dres}
\gamma (\epsilon, T=0) = \frac{3^{1/2}}{2} \frac{1}{(p_Fl)^{3/2}}
\frac{\epsilon^{3/2}}{\epsilon_F^{1/2}} + \frac{\pi}{8}
\frac{\epsilon^2}{\epsilon_F}
\end{equation} 
The first term is exactly Eq. (\ref{tau3D}). It exceeds the second one
for small enough temperatures, $T\tau \ll (p_Fl)^{-2}$, and in the
previous sections it was implicitly supposed that this condition is
satisfied. However, for $T\tau \gg (p_Fl)^2$ the second term in the
most important. This is just the result in the ``clean'' case: the
mean free path does not enter the expressions. So we consider in this
section the case $L \gg l$ for arbitrary temperatures: both diffusive
and clean limits. The result (\ref{3Dres}) is obtained by the diagram
technique and is valid consequently for $\gamma \gg \Delta$ only; in
the diffusive limit this condition is equivalent to $T \gg E_c$, while
in the clean limit for $T = E_c$ one has still $\gamma \gg \Delta$.
The methods used below are valid for $T \ll E_c$, and so in the clean
regime one has some overlap between the results to be obtained and
Eq. (\ref{3Dres}). Below the out-scattering time
$\gamma^{-1}$ for both the GCE and CE cases is calculated.

Prior to the the calculation it is necessary to analyze the 
paper by Sivan, Imry, and Aronov
\cite{Sivan}, devoted to the inelastic scattering rate in the finite
systems. The authors of Ref. \cite{Sivan} calculated exactly the same
quantity as we do --- the scattering rate as a function of energy
$\gamma(\epsilon)$. The methods they used are valid in the
case of ``continuous'' spectrum $\gamma \gg \Delta$ where they have
just reproduced the results (\ref{3Dres}). However, the formal
application of these methods to the opposite case of ``discrete''
spectrum ($\gamma \ll \Delta$) leads to some non-trivial results, that
appear to be not well justified in the approach of
Ref. \cite{Sivan}. Below we present a {\em rigorous} calculation,
valid in the case $\gamma \ll \Delta$, and the results in the range $T
\gg \Delta$ (where level correlations are unimportant) exactly
reproduce those of Ref. \cite{Sivan}. The reasons for this equivalence
are still unclear. Also the effect of level correlations, which can
not be taken into account by the methods of Ref. \cite{Sivan}, is
authomatically incorporated in the approach developed below.  

A general expression for the scattering rate, based on the
perturbation theory of the Coulomb interaction, is
\begin{equation} \label{general}
\gamma_{\lambda} = 2\pi \sum_{\lambda_1,\lambda_2,\lambda_3}
\mid \langle \lambda, \lambda_2 \vert U(\mbox{\bf r}_1, \mbox{\bf r}_2) \vert
\lambda_1, \lambda_3 \rangle \mid^2 \delta(\epsilon_{\lambda} +
\epsilon_{\lambda_2} - \epsilon_{\lambda_1} - \epsilon_{\lambda_3})
f_{\lambda_2} (1 - f_{\lambda_1})(1 - f_{\lambda_3})
\end{equation}
$$\langle \lambda, \lambda_2 \vert U \vert
\lambda_1, \lambda_3 \rangle \equiv \int d\mbox{\bf r}_1 d\mbox{\bf
r}_2 U(\mbox{\bf r}_1, \mbox{\bf r}_2) \psi_{\lambda}^* (\mbox{\bf
r}_1)  \psi_{\lambda_1} (\mbox{\bf r}_1) \psi_{\lambda_2}^*
(\mbox{\bf r}_2) \psi_{\lambda_3} (\mbox{\bf r}_2)$$ 

Here $\vert \lambda_i \rangle \equiv \psi_{\lambda_i} (\mbox{\bf r})$ are
exact single-particle states -- single-electron states in the unique
disorder realization in a quantum 
dot  (note that all four states need to be different) --- 
$f_{\lambda_i}$ are Fermi distribution functions
(the CE case needs a more delicate treatment though; see below) and
$U$ stands for the electron-electron interaction. The expression
(\ref{general}) should be averaged over the disorder realizations. 

\subsection{Coulomb interaction in the restricted geometry}

The first thing is to calculate the screened
Coulomb interaction in the restricted geometry. For the temperatures
(energies) below the Thouless energy one can study the static screening
only. The latter is a
solution to the equation as follows:
\begin{equation} \label{rest}
\nabla^2 U(\mbox{\bf r}_1, \mbox{\bf r}_2) = \kappa^2 \theta(\mbox{\bf
r}_1) \theta(\mbox{\bf r}_2) U + 4\pi e^2 \delta(\mbox{\bf r}_1 -
\mbox{\bf r}_2) 
\end{equation}
$\kappa = (4\pi e^2 \nu)^{1/2}$ being the inverse Debye length, $\nu
\equiv \nu_3$. The functions $\theta (\mbox{\bf r})$ are equal to unity
and zero inside and outside the sample respectively. This equation can
not be solved exactly for a rectangular sample, and one should
introduce some approximations. 

In order to make these approximations clear, we consider first
Eq. (\ref{rest}) in a 
simplest geometry, where it can be solved. Namely, if the
sample occupies a half-space $x > 0$, the exact solution to Eq. (\ref{rest})
in the region $x >0$, $x' > 0$ has the form:
\begin{equation} \label{B2}
U(\mbox{\bf r}_1, \mbox{\bf r}_2) = \int \frac{dq_xdq_y}{(2\pi)^2} 
\exp \left[ iq_y(y-y')+iq_z(z-z') \right]
f_q(x,x'),
\end{equation}
\begin{equation} \label{B3}
f_q(x,x') = -\frac{2\pi e^2}{p} \exp \left( -p\vert x - x' \vert
\right) - \frac{2\pi e^2}{p} \frac{p-q}{p+q} \exp \left[ -p \left ( x
+ x' \right) \right], \ \ \ \ \ p = (q^2 + \kappa^2)^{1/2} 
\end{equation}
Here the first term is essentially the screened Coulomb interaction in
the continuous media; it is translationally invariant and for small
screening length $\kappa^{-1}$ is proportional to $\delta(x -
x')$. The second term is due to the restricted geometry effects; it is
important that it be essentially nonzero when both $x$ and $x'$ lie in
a narrow layer along the boundary of the sample; the thickness of this
layer is of order $\kappa^{-1}$.

Now consider another problem. Let us impose the boundary condition in
Eq. (\ref{rest}): $U \mid_{x=0} = 0$ (and respectively $U$ is nonzero only
when both $x$ and $x'$ lie inside the sample). Then the both
$\theta$-factors are identically equal to unity, and we obtain: 
\begin{equation} \label{B4}
f_q(x,x') = -\frac{2\pi e^2}{p} \exp \left( -p\vert x - x' \vert
\right) + \frac{2\pi e^2}{p} \exp \left[ -p \left ( x + x' \right)
\right] 
\end{equation}

If another boundary condition is imposed: $\partial U/\partial x
\mid_{x=0} = 0$, a solution is
\begin{equation} \label{B5}
f_q(x,x') = -\frac{2\pi e^2}{p} \exp \left( -p\vert x - x' \vert
\right)
\end{equation}

It is seen that the results the problems (\ref{B4}) and (\ref{B5}) yield 
{\em inside} the sample ($x>0,x'>0$) differ from
exact ones (\ref{B3}) by the contribution that is nonzero only near
the boundary 
of the sample. Now one should recollect that the initial problem requires
only the matrix elements of the screened Coulomb potential, and so one has
to consider the region inside the sample only, and the contribution of
the boundary term is small in comparison with the main one by a
factor $(\kappa L)^{-1} \ll 1$. So Eq. (\ref{rest}) may be replaced
by a more simple one, with $\theta$-factors set to be equal to unity and the
boundary condition imposed. Moreover, since the range of the
interaction is $\kappa^{-1}$, i.e. is extremely small, this result is
not sensitive to either boundary conditions or the sample's geometry,
and one may perform this replacement for our rectangular sample as well.

So, turning to the case of the rectangular sample, one easily obtains: 
\begin{equation} \label{B6}
U(\mbox{\bf r}, \mbox{\bf r}') = 4\pi e^2 \sum_q
\frac{\varphi_q(\mbox{\bf r}) \varphi_q(\mbox{\bf r}')}{q^2 +
\kappa^2} 
\end{equation}
Here $\varphi_q(\mbox{\bf r})$ are the eigenfunctions of the Laplace
operator with appropriate boundary conditions; the eigenvalues are
equal to $-q^2$. In the case of the specified cubic geometry one gets
$$\mbox{\bf q} = (\pi/L)(n_x, n_y, n_z), \ \ \ \ n_i = 0,1,2,\dots$$
If the system is charge neutral, the $q = 0$ mode
should be dropped in the summation.

\subsection{Calculation of the matrix element}

Now we return to the equation (\ref{general}). Actually the squared
absolute value of the Coulomb interaction matrix element contains the
product of eight single-particle eigenfunctions, and our statement is
that this combination only weakly depends on the energies of these
states, and therefore can be impurity averaged separately from others,
energy dependent, factors, yielding the constant $U^2$. The results
obtained below follow in fact from the quasi-classical approximation
\cite{GE,Chakr}.   

One needs to average 
\begin{equation} \label{f8}
J(\mbox{\bf r}_1, \mbox{\bf r}_2, \mbox{\bf r}_3, \mbox{\bf r}_4) =
\langle \psi_{\lambda}^*(\mbox{\bf r}_1) \psi_{\lambda_1}(\mbox{\bf
r}_1) \psi_{\lambda}(\mbox{\bf r}_3) \psi_{\lambda_1}^*(\mbox{\bf r}_3)
\psi_{\lambda_2}^*(\mbox{\bf r}_2) \psi_{\lambda_3}(\mbox{\bf
r}_2) \psi_{\lambda_2}(\mbox{\bf r}_4) \psi_{\lambda_3}^*(\mbox{\bf
r}_4) \rangle
\end{equation}
so that
\begin{equation} \label{U21}
U^2 = \int d\mbox{\bf r}_1 d\mbox{\bf r}_2 d\mbox{\bf r}_3 d\mbox{\bf r}_4
U(\mbox{\bf r}_1,\mbox{\bf r}_2) U(\mbox{\bf r}_3, \mbox{\bf r}_4)  
J(\mbox{\bf r}_1, \mbox{\bf r}_2, \mbox{\bf r}_3, \mbox{\bf r}_4)
\end{equation}

In the particular case (\ref{B6}) $U(\mbox{\bf r},\mbox{\bf r}') =
\nu^{-1} \delta(\mbox{\bf r} - \mbox{\bf r}')$ one obtains
\begin{equation} \label{U22}
U^2 = \nu^{-2} \int d\mbox{\bf r}_1 d\mbox{\bf r}_3 
J(\mbox{\bf r}_1, \mbox{\bf r}_1, \mbox{\bf r}_3, \mbox{\bf r}_3)
\end{equation}

The quantity $J$ contains the constant part, that corresponds to the
calculation in the Gaussian ensemble, and coordinate-dependent
contributions of higher modes. 

{\bf Gaussian ensemble}. The constant part can be easily
calculated. It is convenient to introduce the discrete notations. 
If one considers $N$ electrons ($N \sim \epsilon_F/\Delta \gg 1$),
and consequently splits the system over $N$ elementary volumes with
positions $\mbox{\bf r}_i$, then the values of each two eigenfunctions
in each two elementary volumes are in the Gaussian ensemble
independent except for the constraints due to the orthogonality and
normalization conditions for these eigenfunctions. In the leading
order in $N^{-1}$ one has:
\begin{eqnarray} \label{gaens}
& & J_G(\mbox{\bf r}_1, \mbox{\bf r}_1, \mbox{\bf r}_3, \mbox{\bf r}_3)
= \left[
\begin{array}{lr}
V^{-4}, & \mbox{\bf r}_1 = \mbox{\bf r}_3 \\
2N^{-2}V^{-4}, & \mbox{\bf r}_1 \ne \mbox{\bf r}_3 
\end{array} \right.
\end{eqnarray}
$V$ being the volume of the system.

Hence the contribution to $U^2$ from the Gaussian ensemble is
\begin{equation} \label{Ugauss}
U^2_{G} = \nu^{-2} V^{-2} N^{-1}  = \Delta^2/N  = \alpha \Delta^3 /
\epsilon_F 
\end{equation}
Here $\alpha$ is a numerical coefficient. It can be
adjusted from the comparison with the clean limit in the overlap
parameter range (see below); this gives $\alpha = 1/8$. Note
that the 
main contribution to the integral (\ref{U21}) comes from the range
where all four coordinates coincide; in the continuous notation, this
means that the distance between these points is of order of the
screening length.  

{\bf Higher modes.} The contribution of the higher modes is concerned
with the diffusion processes. In particular, the coordinate-dependent
part of the equation (\ref{f8}) 
describes the diffusion of an electron from point $\mbox{\bf
r}_1$ to point $\mbox{\bf r}_3$, and the diffusion of another
electron from point $\mbox{\bf r}_2$ to point $\mbox{\bf
r}_4$. It is reasonable to assume that if point $\mbox{\bf r}_1$ is
far enough from point $\mbox{\bf r}_2$ (in the discrete terms,
these two points lie in different elementary volumes, or, alternatively,
the distance between these points exceeds several interatomic
distances), and point $\mbox{\bf r}_3$ is far enough from 
point $\mbox{\bf r}_4$, this diffusion processes are independent:
\begin{equation} \label{f4}
J_{HM}(\mbox{\bf r}_1, \mbox{\bf r}_2, \mbox{\bf r}_3, \mbox{\bf r}_4) =
\langle \psi_{\lambda}^*(\mbox{\bf r}_1) \psi_{\lambda_1}(\mbox{\bf
r}_1) \psi_{\lambda}(\mbox{\bf r}_3) \psi_{\lambda_1}^*(\mbox{\bf
r}_3) \rangle \langle 
\psi_{\lambda_2}^*(\mbox{\bf r}_2) \psi_{\lambda_3}(\mbox{\bf
r}_2) \psi_{\lambda_2}(\mbox{\bf r}_4) \psi_{\lambda_3}^*(\mbox{\bf
r}_4) \rangle
\end{equation}
If, however, these pairs are close to each other (in particular, this
is the case for the short-ranged interaction), additional contributions
from the diffusion process $\mbox{\bf r}_1 \to \mbox{\bf r}_4$,
$\mbox{\bf r}_2 \to \mbox{\bf r}_3$ appear, and the expression
(\ref{f4}) acquires a coefficient 2. 

The average of four eigenfunctions of the type (\ref{f4}) can be
calculated up to the terms of order $g^{-2}$, with $g = E_c/\Delta$
being the conductance. The result reads \cite{GE,BM}:
\begin{equation} \label{aver}
\langle \psi_{\lambda}^*(\mbox{\bf r}) \psi_{\lambda'}(\mbox{\bf
r}') \psi_{\lambda}(\mbox{\bf r}') \psi_{\lambda'}^*(\mbox{\bf
r}') \rangle = (\Delta/\pi V) \int_0^{\infty} \left( W_{\mbox{\bf r}}
(\mbox{\bf r}', t) - V^{-1} \right) dt 
\end{equation}
Here $W_{\mbox{\bf r}} (\mbox{\bf r}', t)$ is the probability
to find an electron at the time moment $t$ in the point $\mbox{\bf r}'$
if it was in the point $\mbox{\bf r}$ in the time moment $t = 0$. 
This probability obeys the diffusion equation: 
\begin{equation} \label{B9}
\partial W/\partial t = D \nabla^2_{\mbox{\bf r}'} W, \ \ \ W(\mbox{\bf
r}', t=0) = \delta(\mbox{\bf r}' - \mbox{\bf r}) 
\end{equation}
Integrating Eq.(\ref{B9}) over the time variable and taking into account that
for $t \to \infty$ the distribution tends to be uniform one: 
$W = V^{-1}$, one obtains:
\begin{equation} \label{B10}
\int_0^{\infty} W dt = V^{-1} + \sum_{q \ne 0} (Dq^2)^{-1} \varphi_q(\mbox{\bf
r}) \varphi_q(\mbox{\bf r}') 
\end{equation}
Finally, combining Eqs. (\ref{B5}), (\ref{U21}), (\ref{f4}),
(\ref{aver}), and (\ref{B10}), one obtains for the contribution of the
higher modes to the interaction matrix element:
\begin{equation} \label{Ukw}
U^2_{HM} = \frac{2\Xi \Delta^4}{\pi^4 E_c^2} 
\end{equation}
The constant $\Xi$ is given by
$$\Xi = (\pi/L)^4 \sum_{q \ne 0} q^{-4} = 
\sum_{n_x + n_y + n_z > 0} (n_x^2 + n_y^2 + n_z^2)^{-2} \approx 5 $$

\medskip

If one rewrites the contribution of the higher modes as $U^2 \sim
\Delta^2 g^{-2}$, $g = E_c/\Delta$ being the conductance, it is 
easily seen that it has the same contribution as that of the Gaussian
ensemble, except for the factor $N^{-1}$ being replaced with
$g^{-2}$. Consequently one obtains $U^2_G/U^2_{HM} \sim g^2/N$ \cite{foot3}. 
In large sample, $L \gg p_Fl^2$, we obtain $g^2 \ll N$, and
consequently the contribution of higher modes is the leading one. In
the opposite case, $L \ll p_Fl^2$, the Gaussian ensemble produces the
leading term. The latter is universal, i.e. does not contain any information
about the disorder. It is shown below that this contribution leads to
the clean-limit result $\gamma \gg T^2/\epsilon_F$.

To summarize, we have obtained the following expression for the
averaged matrix element of the Coulomb interaction:
\begin{eqnarray} \label{Ufin}
U^2 \equiv \mid \langle \lambda, \lambda_2 \vert U \vert \lambda_1,
\lambda_3 \rangle \mid^2 = \left[
\begin{array}{lrc}
(2\Xi/\pi^2)\Delta^4/E_c^2, & L \gg p_Fl^2 & (g^2 \ll N) \\
\Delta^3/8\epsilon_F, & L \ll p_Fl^2 & (g^2 \gg N)
\end{array}
\right.
\end{eqnarray}

\subsection{Uncorrelated case}

Further calculations are different for GCE and CE. We start from the
most simple GCE case. In this situation the averaged sum over the
three different states $\lambda_1, \lambda_2, \lambda_3$ in Eq. 
(\ref{general}) is essentially the integral over three energies,
corresponding to these states, multiplied by the normalizing factor
$\Delta^{-3}$ and the four-point correlation function $R_4$, that is
responsible for the level repulsion for small energy differences:
\begin{equation} \label{GCE-1}
\gamma(\epsilon) = 2\pi U^2 \Delta^{-3} \int d \epsilon_1
d \epsilon_2 d \epsilon_3 f_2 (1 - f_1) (1 - f_3) \delta(\epsilon +
\epsilon_2 - \epsilon_1 - \epsilon_3) R_4 (\epsilon, \epsilon_1,
\epsilon_2, \epsilon_3)
\end{equation}
Now we set $\epsilon = 0$. As was already mentioned, an analysis
gives the same energy dependence $\gamma(\epsilon, T=0)$ as the
temperature one $\gamma(\epsilon=0, T)$ obtained below, apart from
the numerical coefficients.  

The characteristic scales of variation for the Fermi functions and 
the correlation function $R_4$ are $T$ and $\Delta$ respectively. So
for $T \gg \Delta$ the correlation function can be replaced with its
asymptotic expressions for large values of arguments, i.e. $R_4 = 1$
\cite{matr}. In physical terms this means that the level correlation
does not play any role for the electron-electron scattering rate provided
$T \gg \Delta$. Note also that in this limiting case the GCE and
CE situations coincide essentially, for the number of excited
quasiparticles is large. 

Direct calculation of the integral gives for $T \gg \Delta$:       
\begin{eqnarray} \label{uncorr}
\gamma_{uc} = \frac{\pi^3}{2} \frac{U^2T^2}{\Delta^3} = \left[
\begin{array}{lrc}
(\Xi/\pi)(T^2\Delta/E_c^2) & L \gg p_Fl^2 & (g^2 \ll N) \\
(\pi^3/16)(T^2/\epsilon_F) & L \ll p_Fl^2 & (g^2 \gg N) 
\end{array} \right.
\end{eqnarray}
(Fig.~4). The appearance of the mean level spacing $\Delta$ in the
expression for the case where the level correlation is absent, should
not be misleading: it just stands for a combination $(\nu L)^{-1}$, and
is introduced for convenience.

The upper line of Eq. (\ref{uncorr}) (range III on Fig.~4) corresponds
to the case of ``large'' dots. The dependence $\gamma \propto
T^2$ is universal and is not sensitive to the geometry of the
sample. However, the coefficient depends both of the sample's size and
the mean free path. It is valid for $T \ll E_c$, and gives in this
parameter range $\gamma \ll \Delta$. Hence the spectrum is discrete
and this result can not be obtained in the perturbation
theory. Note, however, that the if one formally applies the
perturbative, e.g. those, developed in \cite{Schmid} or \cite{Sivan},
methods to the range $T \ll E_c$, one obtains the same result $\gamma
\propto T^2\Delta/E_c^2$, $\gamma \ll \Delta$. We have given above a
{\em rigorous, self-consistent} derivation of this result.   

The lower line of Eq. (\ref{uncorr}) is essentially the same result
that appears in the clean limit for bulk 3D system. It does not
contain either the size of the system or the diffusion properties
(such as the mean free path). However, the range of validity for this
result ($L \ll p_Fl^2, \Delta \ll T \ll E_c$) is rather
different. This range includes both cases discrete and continuous
spectra, and consequently we have an overlap between the perturbation
theory and the RMT calculation (region between curves 1 and 3,
Fig.~4). Note that this ``clean'' behavior is observed in small enough
dots even for small temperatures $T\tau \ll (p_Fl)^{-2}$ (cf
\cite{Sivan}). 

\subsection{Effect of level correlation}

For $T \ll \Delta$ the correlation function can be replaced by its
expansion for small arguments; in the particular case of the 
Gaussian Unitary Ensemble (GUE) one has (see Appendix B):
$$R_4 (0, \omega, \omega + \Omega, \Omega) = (\pi/\Delta)^{12}
(212625)^{-1} (\omega^8\Omega^4 - 2\omega^6\Omega^6 +
\omega^4\Omega^8)$$
Straightforward calculation gives:
\begin{equation} \label{GUE-CE}
\gamma_{corr} = \pi^4 \beta \frac{U^2T^2}{\Delta^3} \left( \frac{\pi
T}{\Delta} \right)^{12} = 
2\pi \beta \left( \frac{\pi T}{\Delta}
\right)^{12} \gamma_{uc}; \ \ \ \ \ \beta = \frac{59\pi^{11}}{94500} \sim
1000 
\end{equation}
In particular, one obtains
\begin{eqnarray} \label{qca}
\gamma_{corr} = \left[ \begin{array}{lrc}
\beta \Xi (\Delta T^2/E_c^2)(\pi T/\Delta)^{12}, & L \gg p_Fl^2
& (g^2 \ll N) \\ 
(\pi^4/8)(T^2/\epsilon_F)(\pi T/\Delta)^{12}, & L \ll p_Fl^2
& (g^2 \gg N)  
\end{array} 
\right.
\end{eqnarray}
The twelfth power in the result can be easily explained. One needs to
find 
four energies close to each other; these four energies form six pairs,
and the contribution of each pair is proportional to
$(\omega/\Delta)^2$ in the GUE, $\omega$ being the difference between
the energies in a pair. Consequently one obtains the extra factor
$(T/\Delta)^{12}$ in comparison with the uncorrelated case. In a
similar way, the contribution from each pair is proportional to $\vert
\omega/\Delta \vert$ in the Gaussian Orthogonal Ensemble (GOE) and to
$(\omega/\Delta)^4$ in the Gaussian Symplectic Ensemble (GSE), and so the
results are $\gamma \propto \gamma_{uc}(T/\Delta)^6$ (GOE) and 
$\gamma \propto \gamma_{uc}(T/\Delta)^{24}$ (GSE).

The CE case is more complicated. As we argue above, for $T \gg \Delta$
it yields the same results as the GCE, and so the only interesting
situation is $T \ll \Delta$. In the expression (\ref{general}) the
functions $f$ are still Fermi distribution functions, however the
chemical potential needs now to be pinned to one of the energy levels:
$\mu = \epsilon_l + 0$. This fact changes all results considerably, as
the probability to find $\mu$ in some gap between levels depends no
longer on the width of this gap \cite{Shkl,Lehle}. The
arbitrariness of selection of the pinned level is removed by the
averaging with a weight function that is centered around the ``mean
value'' of the chemical potential $\bar\mu$ and has the support $\delta
\mu$: $\Delta \ll \delta \mu \ll \bar\mu$. If this weight function is
chosen to be a step function, after taking a limit $\delta \mu \to 0$
one obtains:
\begin{equation} \label{CE}
\gamma_{\lambda} = 2\pi U^2 \Delta
\sum_{\lambda_1,\lambda_2,\lambda_3, l} 
\delta(\epsilon_{\lambda} +
\epsilon_{\lambda_2} - \epsilon_{\lambda_1} - \epsilon_{\lambda_3})
\delta (\epsilon_l - \mu + 0) f_{\lambda_2} (1 - f_{\lambda_1})
(1 - f_{\lambda_3})
\end{equation} 

In principle, the summation over $l$ contains terms with
$l=\lambda_i$. These particular terms are reduced after
the disorder averaging to the integrals over three energies with the
four-point correlation function $R_4$. But upon setting $\epsilon = 0$ 
these terms vanish just because of the correlation function,
containing two equal energies ($\epsilon_l = 0$). However for non-zero
$\epsilon$ these terms yield the results:
\begin{equation} \label{fin}
\gamma(\epsilon,T) \sim \left[ \begin{array}{lr}
\gamma_{uc} (\epsilon/T)^4
(T/\Delta)^{12} & \mbox{GUE}\\
\gamma_{uc} (\epsilon/T)^2
(T/\Delta)^{6} & \mbox{GOE}
\end{array} \right.
\end{equation}
All other terms contain the five-point correlation function and hence
can be omitted. Formally for $\epsilon = 0$ one obtains, e.g., in GUE
$\gamma_{CE} \sim \gamma_{uc}(T/\Delta)^{20}$.

\section{Conclusions} \label{ccc}
  
In conclusion, we have investigated the out-scattering time 
$\gamma^{-1}$ appearing due to the electron-electron interaction. Two 
different cases should be distinguished: continuous spectrum ($\gamma
\gg \Delta$), where perturbation theory can be applied, and
non-perturbative case of discrete spectrum ($\gamma \ll \Delta$).

In the case of continuous spectrum the out-scattering time is
essentially the same as the inelastic scattering time 
entering the problems of universal conductance fluctuations and
persistent currents. In 3D and 0D systems it coincides also with 
the phase breaking time, while in 2D and quasi-1D cases these times differ
considerably. In 2D case in some range of parameters we have recovered
the earlier results \cite{AALR,FA,LdS}, but we interpret it as the
out-scattering time rather than the phase-breaking time. Also an
intermediate parameter range between 2D and 0D systems is
investigated. For quasi-1D systems we have obtained principally the results
(\ref{1D})--(\ref{1D-0D}). In purely one-dimensional case for large
enough temperatures $T\tau \gg (p_Fa)^2$ the out-scattering time
occurs to be proportional to $T^{-2/3}$ 
as well as $\tau_{\phi}$, however the former occurs
to be considerably shorter. For lower temperatures $(p_Fa)^{-4} \ll
T\tau \ll (p_Fa)^{-2}$ the out-scattering rate is proportional to
$T^{3/4}$, and becomes larger than the temperature.
Also a transitional region between 1D and
0D system exists, and we have investigated different regimes of the
diffusion. In particular, for $T \gg E_c$ the spectrum is always
continuous, and close enough to $E_c$ we obtain zero-dimensional
behavior: $\gamma \sim \tau_{\phi} \propto T$. Ranges of parameters
corresponding to different regimes are displayed on Fig.~3.

For temperatures below the Thouless energy the spectrum is
discrete. In this case the out-scattering rate coincides with the
phase-breaking one, and it is reasonable to speak just of the
inelastic scattering rate. We have shown that for large enough systems
$L \gg p_F^2l$ ($N \sim \epsilon_F/\Delta \gg g^2$ the latter behaves
itself as $\gamma \sim T^2\Delta/E_c^2$ ($\Delta \ll T \ll E_c$),
while in small systems $N \ll g^2$ the universal dependence $\gamma
\sim T^2/\epsilon_F$ is found. For $T \ll \Delta$ we have obtained an
abundance of results for small -- large / CE -- GCE / GUE -- GOE
cases. In spite of this abundance one should
clearly understand that the inelastic scattering rate due to the
electron-electron collisions is in this parameter range vanishingly
small. So for real systems it is necessary to look for another
mechanisms of inelastic scattering. Electron-phonon scattering seems
to be a good candidate; it has been recently discussed for mesoscopic
systems \cite{Zhou}. Also a coupling to the environment produces an
inelastic rate in the nearly-closed system. 

It is hard to overestimate the help from Professor A.~Schmid, who
initiated this work. Useful discussions with B.~L.~Altshuler,
Y.~Gefen, Y.~Imry, F.~M.~Izrailev, Y.~B.~Levinson, C.~M.~Marcus,
A.~D.~Mirlin, H.~Schoeller, and U.~Sivan are also gratefully
acknowledged. The work was supported 
by the Alexander von Humboldt Foundation  and SFB195 der Deutschen
Forschungsgemeinschaft. 

{\em Note added}. As this paper was being prepared, a number of
related works came to my attention. First, model similar to the
``discrete spectrum'' situation in quantum dots have been recently
studied with the help of the random band matrices, and the results are
rather similar \cite{RBM}. Then, Kamenev and Gefen \cite{KGRen}
studied the role of the external enviroment in the inelastic
broadening, and found this effect to be very strong. They relate this
fact to the phenomenon of the Coulomb blockade. Finally, the recent
unpublished results by Altshuler, Gefen, Kamenev and Levitov
\cite{AGKL} (AGKL) and the comment on their work by Imry \cite{IRen}
are so close to the results obtained above that some discussion is
required. AGKL study the same problem, and reproduce, in fact, the
matrix element (\ref{Ukw}). They interpret it as an overlap between
one-particle $\vert \lambda \rangle$ and three-particle $\vert
\lambda_1, \lambda_2, \lambda_3 \rangle$ states. Consequently this
overlap should be compared with the three-particle level spacing
$\Delta_3 \sim \Delta^3/\epsilon^2$, and this comparison creates a new
energy scale $E^* = (E_c\Delta)^{1/2}$. For $E > E^*$ the
three-particle states are well mixed by the Coulomb interaction, and
the broadened peaks (which are resolved for $E < E_c$) are essentially
a mixture of many-particle states. On the other hand, for $E < E^*$
the single-particle state is mixed with one three-particle state, one
five-particle state, and so on. In this sense the inelastic scattering
rate is zero: the state does not decay at all. AGKL describe this
situation as an analog of the localization transition on the Bethe
lattice. I do not want to address this problem here; however, the
results obtained above for $E < E^*$ can be interpreted rather as the
width of the ``envelope'', formed by the many-particle states around
the single-particle one. I am indebted to the authors of all these
papers for the possibility to become acquainted with their results
prior to publication.

\section*{Appendix A. Dephasing and Q1D - 0D crossover}

Below results for the phase-breaking time in the range of
parameters intermediate between quasi-one- and zero-dimensional case
are obtained. In
principle, one has to perform calculations similar to those of
Ref. \cite{AAK} in restricted geometry, and this does not look quite
hopeful. However, as we are interesting
in the result up to the numerical factor only, it is reasonable to use
the method developed by Stern, Aharonov, and Imry \cite{Imry,Imry1}
that was later applied to a calculation of the quasiparticle
lifetime in a quantum dot \cite{Sivan}. In this approach the phase
uncertainty $P(t_0)$ accumulating by the electron due to the interaction with
the environment (in our particular case it interacts with the
electromagnetic fluctuations) is calculated; the time $t_0$ when this
phase uncertainty becomes of order unity (we will set it exactly 
unity) is associated with the electron lifetime (phase-breaking time).
This phase uncertainty is given by \cite{Sivan}
$$P(t_0) = \frac{2}{La^2} \int_0^{t_0} dt dt' \int_{-\infty}^{\infty}
d\omega \coth \frac{\omega}{2T} \sum_{\mbox{\bf q} \ne 0} \frac{4e^2}{q^2}
\mbox{Im} \left( \frac{1}{\varepsilon (\mbox{\bf q},\omega)} \right)
e^{i\omega(t - t')} \left(S_1 + S_2 + S_3 + S_4 \right)
\eqno(A1)$$
with
$$S_1 = \exp(i\mbox{\bf q}[x_1(t) - x_1(t')]), \ \ S_2 = \exp(i\mbox{\bf
q}[x_2(t) - x_2(t')]),$$
$$S_3 = \exp(i\mbox{\bf q}[x_1(t) -
x_2(t')]), \ \ S_4 = \exp(i\mbox{\bf q}[x_2(t) - x_1(t')])
\eqno(A2)$$
In these expressions (to be the direct generalization of those of
Ref. \cite{Sivan} for the case of finite temperatures) $x_1(t)$ and
$x_2(t)$ are arbitrary (quasiclassical) electron paths, and the 
averaging over
these paths is supposed; $\varepsilon(\mbox{\bf q}, \omega)$ is the
dielectric susceptibility. The expression (A1) is derived for an
infinite system, however, as it varies on scales of order 
of the elastic mean free path $l \ll a$, this expression is
valid in the diffusive regime, and the bulk results may be substituted
for the dielectric susceptibility. It is also important that the
$\mbox{\bf q} = 0$ mode does not contribute to the sum (A1): it is
absent in the open systems while in the closed ones the corresponding
contribution 
is forbidden by the charge neutrality. So the difference between open
and closed systems is not quite important, and the system is assumed to be
closed: $\mbox{\bf q} = \pi (n_x/L, n_y/a, n_z/a)$, $n_x + n_y + n_z >
0$. The results for the open systems depend on boundary conditions,
but in all cases differ by a 
numerical factor of order unity only. 

After the averaging over paths is carried out in exactly the same way
as in Ref. \cite{Sivan}, one obtains:
$$P(t_0) = \frac{48 T}{La^2\pi p_F^2l} \frac{D}{E_c} \sum_{n_x + n_y + n_z
> 0} \frac{1}{n_x^2 + (L/a)^2(n_y^2 + n_z^2)} \int_0^{t_0} dt^+
\int_0^{t^+} \frac{dt}{t} \sin(2Tt) \times$$
$$\times\exp \left[ -\pi^2 E_c t
\left(n_x^2 + (L/a)^2(n_y^2 + n_z^2\right) \right] \eqno(A3)$$
Here $D$ is the three-dimensional diffusion coefficient and we have
taken into account that the main contribution to the integral over
frequencies comes from the region $\vert \omega \vert < 2T$. The
summation is restricted by the condition $q \ll l$. For $D/a^2 \gg T$
(this condition excludes 3D situation) only the values $n_y = n_z = 0$
are important in the summation (A3).  

In the limiting case $E_c \gg t_0^{-1}$ (the inverse situation
corresponds to the ``true'' quasi-1D case and the result is given by
Eq. (\ref{taufi-1D})) the integral can be easily calculated. One gets:
$$P(t_0) = \frac{48 T}{\pi p_F^2lLa^2} \frac{D}{E_c} \sum_{n=1}^{\sim L/l}
\frac{1}{n^2} \left[ t_0 \arctan \frac{2T}{\pi^2 E_c n^2} -
\frac{2T}{4T^2 + \pi^2 E_c^2n^2} \right] \eqno(A4)$$

The phase-breaking time $\tau_{\phi}$ is defined as the time $t_0$
when the phase uncertainty $P(t_0)$ is equal to unity. 
 
The case $E_c \gg T$ corresponds to the ``true'' 0D situation. One has
$$\frac{1}{\tau_{\phi}} = \frac{96 T}{\pi^3 p_F^2 l La^2} \frac{D}{E_c^2}
\sum_{n=1}^{\infty} \frac{1}{n^4} = \frac{5\pi T^2 \tau}{16 p_F^2 a^2}
\left( \frac{L}{l} \right)^3 \eqno(A5)$$ 
This is the result by Sivan, Imry and Aronov \cite{Sivan} up to the
numerical factor. It is seen, however, that in this region the spectrum
occurs to be discrete: $\tau_{\phi}^{-1} \ll \Delta$. Thus, the 0D
region is not subject to our analysis and should be treated by another
methods. (See, however, the discussion after Eq. (\ref{uncorr})).

The case $\tau_{\phi}^{-1} \ll E_c \ll T$ is intermediate between quasi-1D
and 0D regimes. Only terms with $E_cn^2 \ll T$ are important in the
summation (A4), however due to the condition $T \gg E_c$ the summation
can be extended to the infinity. One obtains:
$$\frac{1}{\tau_{\phi}} = \frac{4\pi^2 T L}{(p_Fa)^2 l} \eqno(A6)$$
that is a new result. In the whole region $\tau_{\phi} \ll E_c \ll T$
the spectrum turns out to be continuous and the result (A6) is valid. 

\newpage

\section*{Appendix B. Four-point correlation function in GUE}

Below we follow the generalities given in Refs. \cite{matr}. As an
explicit expression for the four-point correlation is not given
anywhere in literature to the best of our knowledge, we derive it for
the most simple GUE case. It is convenient to use the dimensionless
energies $x = \pi \epsilon/\Delta$ throughout this Appendix.

The first step is to define the functions 
$$Y_i(x_1,x_2,\dots,x_i) = \sum_P s_{12} s_{23} \dots s_{i1}
\eqno(B1)$$
Here 
$$s_{ij} \equiv s(\vert x_i - x_j \vert),\ \ \ \ \ s (x) \equiv
\frac{\sin x}{x}, \eqno(B2)$$
and the summation is carried out over $(i-1)!$ different permutations
of indices. Hence,
$$Y_1 (x)= 1;\ \ \ Y_2 (x_1,x_2) = s^2_{12}; \ \ \ Y_3(x_1,x_2,x_3) =
2s_{12}s_{23}s_{31};$$
$$Y_4 (x_1,x_2,x_3,x_4) = 2s_{12}s_{23}s_{34}s_{41} +
2s_{13}s_{34}s_{42}s_{21} +  2s_{14}s_{42}s_{23}s_{31}$$

Now the correlation functions $R_i(x_1,\dots, x_i)$ can be expressed
as
$$R_1 = 1;\ \ \ R_2(x_1,x_2) = -Y_2(x_1,x_2) + R_1(x_1)R_2(x_2);$$
$$R_3 (x_1,x_2,x_3) = Y_3(x_1,x_2,x_3) + R_1(x_1) R_2(x_2,x_3) +
R_1(x_2) R_2(x_1,x_3) + $$
$$+ R_1(x_3) R_2(x_1,x_2) - 2R_1(x_1)R_1(x_2)R_1(x_3);$$
\medskip
$$R_4(x_1,x_2,x_3,x_4) = -Y_4(x_1,x_2,x_3,x_4) +
\{R_1(x_1)R_3(x_2,x_3,x_4) + R_1(x_2)R_3(x_1,x_3,x_4) +$$
$$+  R_1(x_3)R_3(x_1,x_2,x_4) + R_1(x_4)R_3(x_1,x_2,x_3) \} + \{
R_2(x_1,x_2) R_2(x_3,x_4) + R_2(x_1,x_3) R_2(x_2,x_4) +$$    
$$+ R_2(x_1,x_4) r_2(x_2,x_3) \} - 2\{ R_2(x_1,x_2)R_1(x_3)R_1(x_4) +
R_2(x_1,x_3)R_1(x_2)R_1(x_4) + $$
$$+ R_2(x_1,x_4)R_1(x_2)R_1(x_3) + R_2(x_2,x_3)R_1(x_1)R_1(x_4) +
R_2(x_2,x_4)R_1(x_1)R_1(x_3) +$$
$$+ R_2(x_3,x_4)R_1(x_1)R_1(x_2) \} +
6R_1(x_1)R_1(x_2)R_1(x_3)R_1(x_4)$$

After some algebra one obtains an explicit expression for the
correlation function $R_4$:
$$R_4(x_1,x_2,x_3,x_4) = 1 - 2\{s_{12}s_{23}s_{34}s_{41} +
s_{13}s_{34}s_{42}s_{21} + s_{14}s_{42}s_{23}s_{31} \} + $$
$$\{ s_{12}^2s_{34}^2 + s_{13}^2s_{24}^2 + s_{14}^2s_{23}^2 \} + 2\{
s_{12}s_{23}s_{31} + s_{12}s_{24}s_{41} + s_{13}s_{34}s_{41} +
s_{23}s_{34}s_{42} \} - $$
$$- \{ s_{12}^2 + s_{13}^2 + s_{14}^2 + s_{23}^2 + s_{24}^2 + s_{34}^2
\} \eqno(B3)$$

In our particular case $x_1 + x_3 = x_2 + x_4$, and, taking into
account that the correlation function depends on three differences of
arguments only, one arrives to an expression $R_4(0,x,x+y,y)$. From (B3)
after cumbersome calculations one obtains an expansion of $R_4$ for
$x,y \ll 1$:
$$R_4(0,x,x+y,y) = \frac{1}{212625} \left( x^4y^8 - 2x^6y^6 + x^8y^4
\right) \eqno(B4)$$
The 12th power can be easily explained. Pair correlation function
for small arguments is proportional in GUE $R_2(x) \propto x^2$. As one
has in $R_4$ six pairs of arguments close to each other and each pair
produces the second power, the total expansion starts from the power
twelve. In GOE $R_2(x) \propto x$, and we may easily conclude that the
analogous expansion starts from the sixth power.

As could be expected, for a large value of the arguments $x,y \gg 1$ all
functions $S_{ij}$ are small, and $R_4 = 1$. This fact means just that
the levels are uncorrelated on distances exceeding the mean level
spacing $\Delta$.

\newpage
{\large FIGURE CAPTIONS}

Fig.~1. The diagram representation of the diffusion propagator.

Fig.~2. The diagrams contributing to the interaction block (crossed
rectangle on Fig.~1) for the UCF diffusion propagator. Wavy lines
and dashed lines with a cross denote the electron-electron
interaction and bare impurity scattering $(\pi\nu\tau)^{-1}$ respectively.

Fig.~3. Different regimes for 1D -- 0D crossover. Curves: 1: $g =
E_c/\Delta = 1$; 2: $T\tau = (l/L)^3(p_Fa)^4$ (or $\gamma =
E_c(p_Fa)^2$); 3: $T \tau = (l/L)^2 (p_Fa)^2$ (or $E_c = T(p_Fa)^{-2}$);
4: $T\tau = (l/L)^3(p_Fa)^2$ (or $E_c\tau_{\phi} = 1$); 5: $T\tau =
(l/L)^2$ (or $T = E_c$. Regimes I -- V are described in Sec. {1DD}. 

Fig.~4. 3D -- 0D crossover. Curves: 1: $T = E_c$; 2: $T =
\Delta$. Line 3 ($\gamma = \Delta$) separates the cases of continuous
and discrete spectrum for $L \gg p^Fl^2$. Regimes: I -- clean limit, 
$\gamma \sim T^2/\epsilon_F$; II - bulk diffusive limit, $\gamma
\propto T^{3/2}$; III - 
zero-dimensional diffusive limit; in the ranges IV and V the level
correlation is important.


\begin{references}
\bibitem{foot1} The third important inelastic time is the energy
relaxation time \cite{AA}. It is not discussed in this paper.
\bibitem{foot2} This is done throughout
the whole paper, except for the basic expressions. It should be noted,
however, that in the case of zero-dimensional systems 
the frequency dependence of the out-scattering rate
$\gamma(\epsilon, T=0)$ coincides with its temperature dependence 
$\gamma(\epsilon=0,T)$ up to the numerical coefficient.
\end{references}
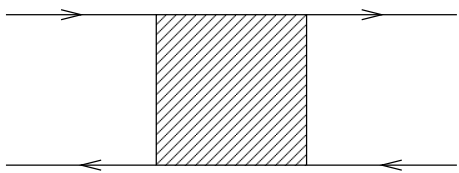
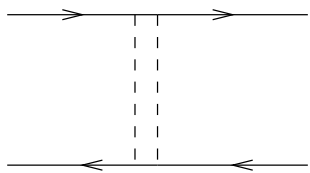
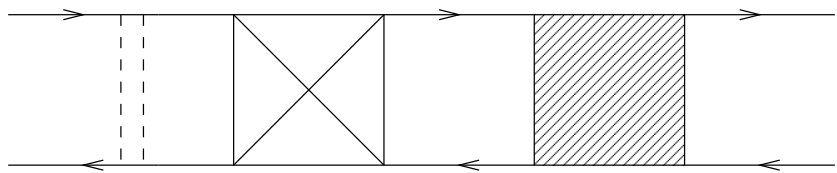
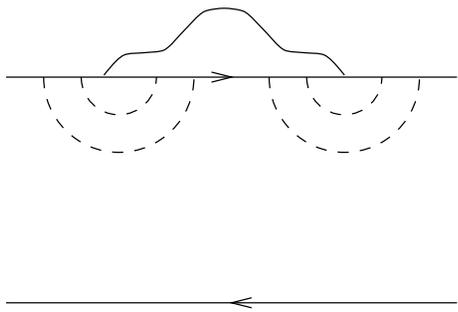
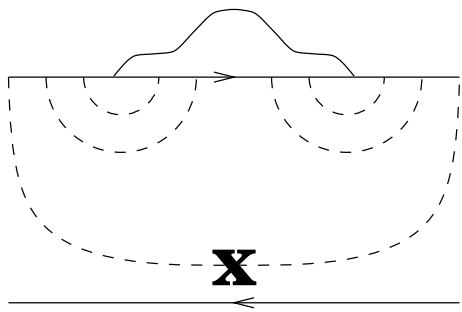
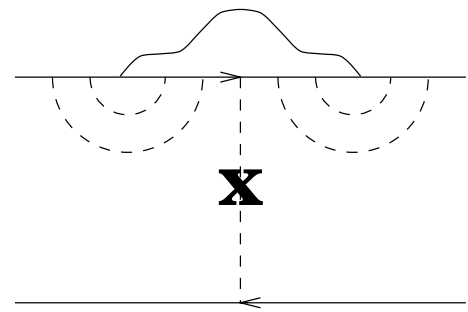
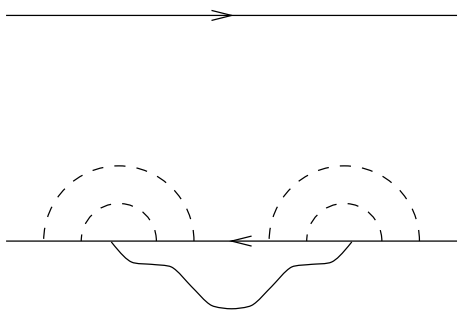
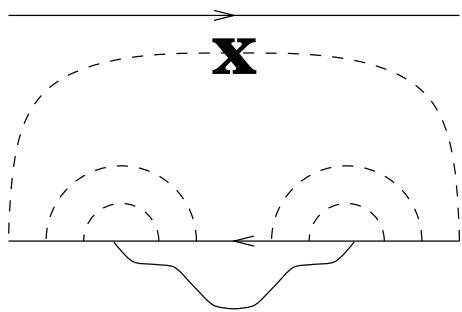
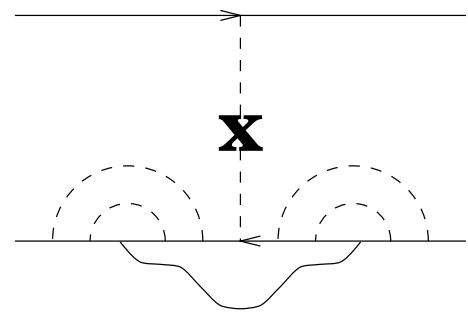
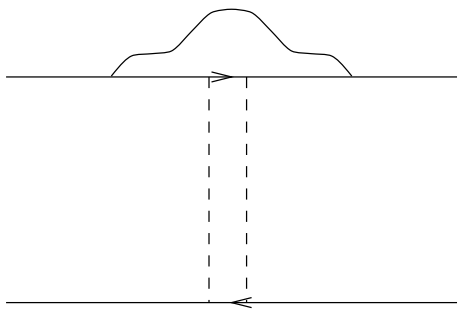
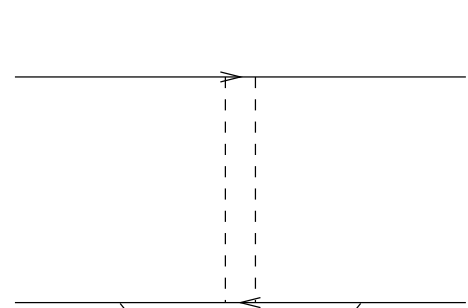
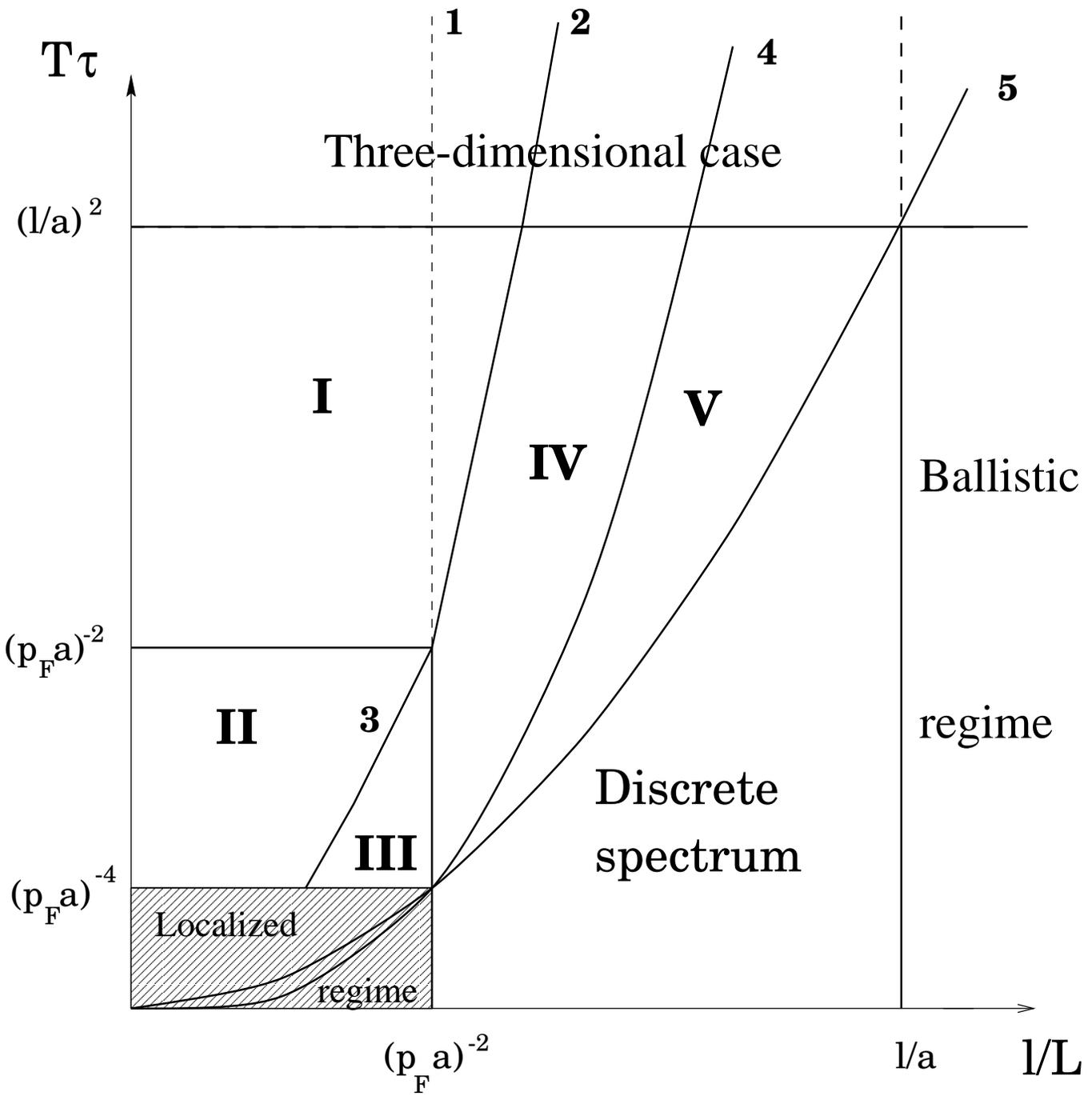
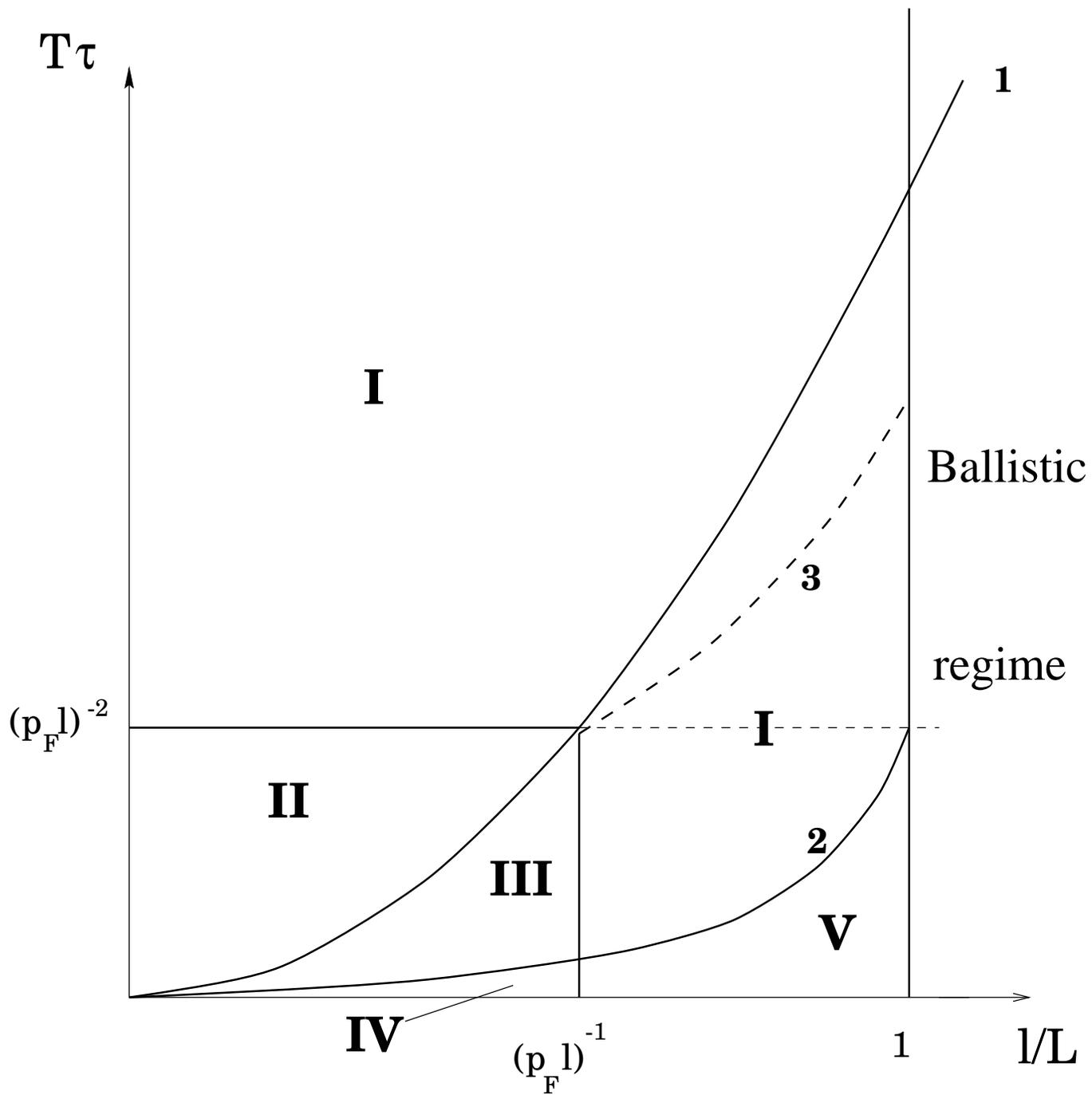
\end{document}